# Выявляемость новой коронавирусной инфекции и смертность от новой коронавирусной инфекции в разных субъектах Российской Федерации

Эдвард Гольдштейн (Edward Goldstein)[1,*]

1. Гарвардская школа общественного здравоохранения, Бостон, США
*. Электронная почта: egoldste@hsph.harvard.edu

## Аннотация

*Актуальность:* Лабораторная диагностика новой коронавирусной инфекции в совокупности с отслеживанием/карантином для контактных лиц является эффективным способом для уменьшения распространения новой коронавирусной инфекции и снижения уровня соответствующей смертности. При этом, практика тестирования на новую коронавирусную инфекцию различается в разных регионах Российской Федерации. Например, в г. Санкт-Петербурге, где самый высокий уровень смертности от COVID-19 на 100,000 человек в Российской Федерации на 25/10/2020, на одну смерть от COVID-19 приходится 15.7 выявленных случаев новой коронавирусной инфекции, а в среднем по Российской Федерации этот показатель составляет 58.1 выявленных случаев новой коронавирусной инфекции, что говорит об ограниченном выявлении новой коронавирусной инфекцию при легких и средних случаях заболевания в Санкт-Петербурге. Также, в ряде других субъектов Российской Федерации, только определенные категории лиц, обращающихся за медицинской помощью с симптомами ОРВИ тестируются на новую коронавирусную инфекцию.
*Материалы и методы:* При более активном тестировании на новую коронавирусную инфекцию в населении, выявляемость новой коронавирусной инфекции (т.е. процент выявленных случаев COVID-19 среди всех случаев заражения новой коронавирусной инфекцией в населении) увеличивается, а коэффициент летальности (процент смертельных случаев среди всех выявленных случаев COVID-19 в населении) уменьшается (т.к. при более активном тестировании, выявляются больше случаев заболевания COVID-19 в легкой и средней форме). Мы используем данные Роспотребнадзора о количестве выявленных случаев новой коронавирусной инфекции и количестве смертей от новой коронавирусной инфекции в разных субъектах Российской Федерации для того, чтобы оценить корреляцию между коэффициентом летальности и уровнем смертности от COVID-19 в разных субъектах Российской Федерации.

*Результаты:* Корреляция между коэффициентом летальности на 25/10/2020 и уровнем смертности от COVID-19 на 100,000 человек на 25/10/2020 в разных субъектах Российской Федерации равна 0.64 (0.50,0.75). В ряде субъектов Российской Федерации, выявляемость новой коронавирусной инфекции относительно низкая, а уровень смертности от COVID-19 относительно высокий.

*Выводы:* Выявляемость новой коронавирусной инфекции является одним из факторов, которые влияют на уровень смертности от COVID-19 в России -- более высокая выявляемость приводит к понижению уровня смертности от COVID-19. Для повышения выявляемости следует тестировать на новую коронавирусную инфекцию всех лиц, обращающихся за медицинской помощью с симптомами ОРВИ, а также принимать дополнительные меры для повышения уровня тестирования на новую коронавирусную инфекцию. Такие меры, в совокупности с карантином для инфицированных и контактных лиц способствует понижению уровня заболеваемости и смертности от COVID-19.

## Abstract

*Relevance:* Laboratory diagnosis of the novel coronavirus (SARS-CoV-2) infection combined with quarantine for contacts of infected individuals affects the spread of the

SARS-CoV-2 infection and levels of related mortality. At the same time, detectability of the SARS-CoV-2 infection (i.e. the proportion of detected COVID-19 cases among all cases of SARS-CoV-2 infection in the population) varies in the different regions of the Russian Federation. For example, in the city of St. Petersburg, where mortality rate for COVID-19 is the highest in the Russian Federation on Oct. 1, 2020, every death for COVID-19 corresponds to 15.7 detected cases of COVID-19 in the population, while the corresponding number for the whole of Russia is 58.1, suggesting limited detection of mild and moderate cases of COVID-19 in St. Petersburg. In particular, in St. Petersburg, as well as in a number of other regions in Russia, only certain categories of individuals presenting for medical care with respiratory symptoms are tested for the SARS-CoV-2 infection.

*Material & Methods:* More active testing for SARS-CoV-2 in the population results in increased detectability (i.e. the proportion of detected COVID-19 cases among all cases of SARS-CoV-2 infection in the population) and decreased case-fatality ratio (CFR, the proportion of deaths among reported COVID-19 cases in the population) – this because under more active testing, the number of mild and moderate cases of COVID-19 increases. We used data from the Russian Federal Service for Surveillance on Consumer Rights Protection and Human Wellbeing (Rospotrebnadzor) on the number of detected cases and the number of deaths from COVID-19 in the different regions of the Russian Federation to examine the correlation between case-fatality ratios and rates of mortality for COVID-19 in different regions of the Russian Federation.

*Results:* The correlation between case-fatality ratios and rates of mortality for COVID-19 in different regions of the Russian Federation on Oct. 25, 2020 is 0.64 (0.50,0.75). For several regions of the Russian Federation, detectability of SARS-CoV-2 infection is relatively low, while rates of mortality for COVID-19 are relatively high.

*Conclusions:* Detectability of the SARS-CoV-2 infection is one of the factors that affects the levels of mortality from COVID-19 – higher detectability contributes to lower rates of mortality from COVID-19. To increase detectability, one ought to test all individuals with respiratory symptoms seeking medical care for SARS-CoV-2 infection, and to undertake additional measures to increase the volume of testing for SARS-CoV-2. Such measures, in combination with quarantine for infected cases and their close contacts help to mitigate the spread of the SARS-CoV-2 infection and diminish the related mortality.

**Введение**

Лабораторная диагностика новой коронавирусной инфекции в совокупности с отслеживанием/карантином для лиц, контактировавших с инфицированным лицом (контактных лиц) является эффективным способом для уменьшения распространения новой коронавирусной инфекции и снижения уровня соответствующей смертности. Например, в Исландии соответствующая диагностика и карантин/самоизоляция активно практикуются – так, используя серологические данные и данные о лабораторной (ПЦР) диагностике новой коронавирусной инфекции, исследователи оценили, что 56% всех случаев новой коронавирусной инфекции в Исландии были диагностированы [1]. Отметим также, что уровень смертности от COVID-19 в Исландии является одним из самых низких в Европе. В австралийском штате Новый Южный Уэльс, распространение новой коронавирусной инфекции в школах весной 2020-го года было очень ограниченно [2]; при этом, в школах проводилось активное тестирование на новую коронавирусную инфекцию, и лиц, контактировавших с зараженными школьниками или персоналом школы отправляли на двухнедельный карантин. В Российской Федерации тоже проводится активное тестирование на новую коронавирусную инфекцию [3], что способствует уменьшению распространения инфекции. Вместе с тем, практика тестирования на новую коронавирусную инфекцию различается в разных регионах Российской Федерации. Например, в ряде регионов Российской Федерации, все лица, обращающиеся за медицинской помощью с симптомами ОРВИ тестируются на новую коронавирусную инфекцию [4,5]; в ряде других регионов Российской

Федерации, только определенные категории лиц (люди старше 65-и лет, медицинские работники, и др.), обращающиеся за медицинской помощью с симптомами ОРВИ тестируются на новую коронавирусную инфекцию [6,7]. Влияние различий в практике тестирования на новую коронавирусную инфекцию на распространение и уровень смертности от COVID-19 в разных регионах Российской Федерации мало изучена.

Влияние тестирования на распространение новой коронавирусной инфекции зависит от *выявляемости* новой коронавирусной инфекции (т.е. процента выявленных случаев COVID-19 среди всех случаев заражения новой коронавирусной инфекцией в населении). При более высокой выявляемости, лабораторно подтверждается большее количество случаев инфицирования в населении, что способствует предотвращению большего количества новых инфекций путем карантина для выявленных случаев и их контактных лиц, что уменьшает темп распространения инфекции в населении. При более активном тестировании на новую коронавирусную инфекцию в населении, выявляемость новой коронавирусной инфекции увеличивается, а коэффициент летальности (процент смертельных случаев среди всех выявленных случаев COVID-19 в населении) уменьшается (т.к. при более активном тестировании, выявляются больше случаев заболевания COVID-19 в легкой и средней форме). Поэтому более низкий коэффициент летальности соответствует более высокой выявляемости новой коронавирусной инфекции в разных регионах Российской Федерации. В этой работе, мы используем данные Роспотребнадзора о количестве выявленных случаев новой коронавирусной инфекции и количестве смертей от новой коронавирусной инфекции в разных субъектах Российской Федерации для того, чтобы оценить связь между выявляемостью, коэффициентом летальности и уровнем смертности от COVID-19 в разных субъектах Российской Федерации.

***Цель исследования****:* Оценка связи между *выявляемостью* (т.е. процентом выявленных случаев COVID-19 среди всех случаев заражения новой коронавирусной инфекцией в населении), коэффициентом летальности (процентом смертных случаев среди всех выявленных случаев COVID-19 в населении) и уровнем смертности от новой коронавирусной инфекции на 100,000 человек в разных субъектах Российской Федерации.

**Материалы и Методы**

*Данные*

Мы использовали данные Роспотребнадзора о количестве выявленных случаев новой коронавирусной инфекции и количестве смертей от новой коронавирусной инфекции в разных субъектах Российской Федерации на 25/10/2020 [8]. Мы также использовали данные Росстата о численности населения в разных субъектах Российской Федерации на 1-го января, 2020 [9] в целях оценки уровня смертности от новой коронавирусной инфекции на 100,000 человек в разных субъектах Российской Федерации.

*Статистический анализ*

Мы оценили корреляцию между коэффициентом летальности (процентом смертных случаев среди всех выявленных случаев новой коронавирусной инфекции в населении) и уровнем смертности от новой коронавирусной инфекции на 100,000 человек в разных субъектах Российской Федерации на 25/10/2020.

**Результаты**

На Рис. 1 изображены коэффициенты летальности (процент смертельных случаев среди всех выявленных случаев новой коронавирусной инфекции в населении) и уровни смертности от новой коронавирусной инфекции на 100,000 человек в разных субъектах Российской Федерации на 25/10/2020. Корреляция между коэффициентом летальности и уровнем смертности от новой коронавирусной инфекции в разных субъектах Российской Федерации на 25/10/2020 равна 0.64 (0.50,0.75). Среди всех субъектов Российской Федерации, в г. Санкт-Петербурге (правый верхний угол на Рис. 1) самый высокий уровень смертности от COVID-19 на 100,000 человек и самый высокий коэффициент летальности -- 6.4% (самая низкая выявляемость новой коронавирусной инфекции). На 25/10/2020, на одну смерть от COVID-19 в Санкт-Петербурге приходится 15.7 выявленных случаев новой коронавирусной инфекции, а в среднем по Российской Федерации этот показатель составляет 58.1 выявленных случаев новой коронавирусной инфекции, что говорит об ограниченном выявлении новой коронавирусной инфекцию при легких и средних случаях заболевания в Санкт-Петербурге.

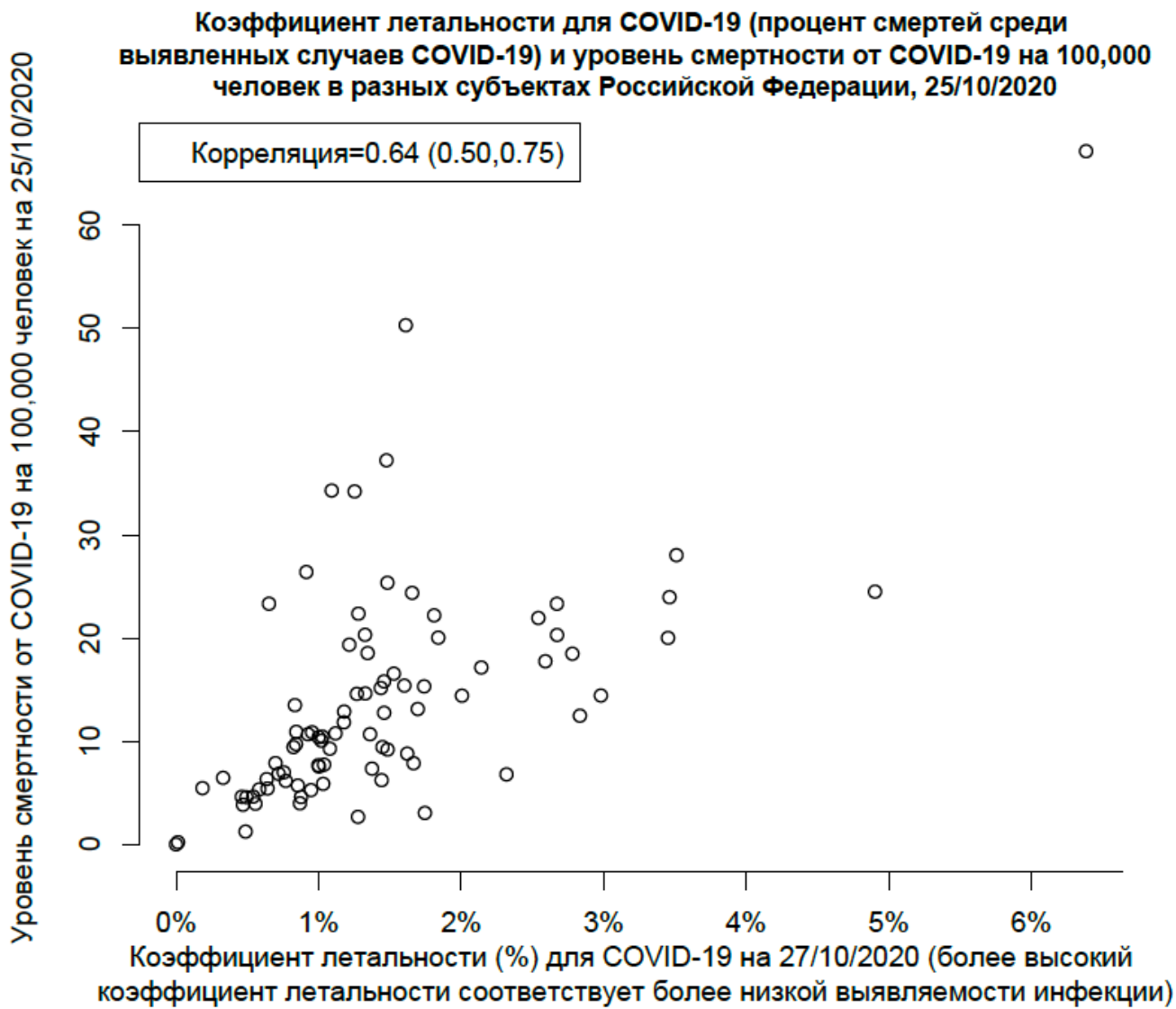

**Рис. 1:** Коэффициенты летальности (процент смертельных случаев среди всех выявленных случаев новой коронавирусной инфекции в населении) и уровни смертности от новой коронавирусной инфекции на 100,000 человек в разных субъектах Российской Федерации на 25/10/2020.

**Выводы**

Активное тестирование на новую коронавирусную инфекцию с последующим карантином для инфицированных и контактных лиц способствует уменьшению распространения новой коронавирусной инфекции [10] и снижению уровня соответствующей смертности. Высокий уровень тестирования и выявления новой коронавирусной инфекцию наблюдается в ряде стран [1,11], включая Российскую Федерацию. Это приводит к высокому уровню *выявляемости* новой коронавирусной инфекции (т.е. проценту выявленных случаев COVID-19 среди всех случаев заражения новой коронавирусной инфекцией в населении) – так, в Исландии выявляемость новой коронавирусной инфекции оценена в 56% [1]. В Российской Федерации, критерии и практика тестирования на новую коронавирусную инфекцию различаются в разных регионах. В частности, в ряде субъектов Российской Федерации практикуется тестирование на новую коронавирусную инфекцию для всех людей с респираторными симптомами, которые обращаются за медицинской помощью [4,5]; в ряде других субъектов Российской Федерации, в амбулаторных условиях тестируют только определённые категории лиц (людей старше 65-и лет, медицинских работников, и др.) [6,7]. При этом, влияние различий в практике тестирования на новую коронавирусную инфекцию на уровень заболеваемости и смертности от новой коронавирусной инфекции в разных регионах Российской Федерации мало изучено.

В этой работе, используя данные Роспотребнадзора о количестве выявленных случаев новой коронавирусной инфекции и количестве смертей от новой коронавирусной инфекции, мы установили высокую корреляцию между коэффициентом летальности (процентом смертельных случаев среди всех выявленных случаев новой коронавирусной инфекции в населении) и уровнем смертности от новой коронавирусной инфекции в разных субъектах Российской Федерации. Отметим также, что более низкий коэффициент летальности соответствует более высокой выявляемости новой коронавирусной инфекции. Соответственно, тестирование на новую коронавирусную инфекцию и выявляемость новой коронавирусной инфекции являются одними из факторов, которые влияют на смертность от новой коронавирусной инфекции в России. В частности, более высокая выявляемость новой коронавирусной инфекции приводит к понижению уровня смертности от новой коронавирусной инфекции. Для повышения выявляемости следует тестировать на новую коронавирусную инфекцию всех лиц, обращающихся за медицинской помощью с симптомами ОРВИ, а также принимать дополнительные меры для повышения уровня тестирования на новую коронавирусную инфекцию. Такие меры, в совокупности с карантином для инфицированных и контактных лиц способствует понижению уровня заболеваемости и смертности от COVID-19.

**Список литературы**